\documentclass[12pt]{article}

\usepackage[letterpaper, portrait, margin=1in]{geometry}
\usepackage[utf8]{inputenc}
\usepackage{float}
\usepackage{authblk}

\usepackage{caption}
\usepackage[style=nature]{biblatex}
\usepackage{mathtools}
\usepackage{amssymb}
\usepackage[group-separator={,}]{siunitx}
\usepackage{physics}

\title{Non-Ellipsoidal Gravity-Based Definitions of \\Planetary Surface Area and Other Geodetic Measures}
\author[]{Kai Xu}
\affil[]{Yale University}
\date{}

\begin{document}

\graphicspath{{./figures/}}

\maketitle

\begin{abstract}

This paper introduces new definitions of common geodetic measures on a planetary surface (namely surface area, path length, and mean value or other statistical parameters of a surface function) that are not based on a datum such as a reference ellipsoid. Instead, the so-called datumless geodetic measures are based on physically meaningful formulations that rely only on the actual planetary surface and gravity. The datumless measures provide universally standardized measurements on any terrestrial object, including non-ellipsoidal asteroids and comets. Conveniently, on fairly round planets such as Earth and Mars, the datumless measures yield very similar values as corresponding geodetic measures on a reference ellipsoid. Like their ellipsoidal counterparts, the datumless measures quantify area and length in the familiar ``bird's-eye view'' or ``horizontal, normal-to-gravity'' sense. Far from being purely theoretical, the datumless measures can be approximated in GIS software using a digital elevation model and a gravity model such as a geoid.

\end{abstract}

\section{Introduction}

Since the surface of a planet\footnote{For conciseness, a \textit{planet} will refer to any terrestrial object, including moons, asteroids, and comets.} is fractal-like, geodetic measures on the planetary surface (such as surface area or path length) have thus far relied on approximating the planet's shape with a reference ellipsoid, also known as a datum \parencite{ellipsoid-area, geodesic-distance}. So for instance, the surface area of the United States would refer to the surface area of a corresponding region on an ellipsoid.

Nevertheless, reference ellipsoids are still arbitrary to an extent, as planets are not perfect ellipsoids. On many irregularly-shaped asteroids and comets (such as Comet 67P), reference ellipsoids provide an inherently crude representation of surface area and other geodetic measures \parencite{comet-67p}. Even on roughly ellipsoidal planets such as Earth, disagreements on how exactly to define the reference ellipsoid have resulted in many ellipsoids being defined over the years, with geodetic measurements differing depending on the ellipsoid used \parencite{geodesy}.

In light of these factors, this paper introduces new definitions of common geodetic measures (namely surface area, path length, and statistical parameters on a surface or path) that do not rely on a datum such as a reference ellipsoid. Known as the datumless geodetic measures, these definitions are based on physically meaningful formulations that rely only on the actual planetary surface and gravity. The datumless measures provide universally standardized measurements that are applicable to any terrestrial object, even non-ellipsoidal ones. Conveniently, on fairly round objects, the datumless measures yield very similar values as corresponding measures on a reference ellipsoid. Like the ellipsoidal measures, the datumless measures quantify area and length in the familiar ``bird's-eye view'' or ``horizontal, normal-to-gravity'' sense. The datumless measures can be sufficiently approximated in GIS software using a digital elevation model and a gravity model such as a geoid.

\section{Defining the Planetary Surface}

A planet's gravitational field can be represented by constant surfaces of gravitational potential surrounding the planet, also known as equipotential surfaces.\footnote{In physics, gravity refers to the fundamental force of gravitation, whereas in geophysics, gravity refers to the sum of gravitation and the centrifugal force. Since the datumless measures can be represented under either definition with usually minor resulting differences, the reader can feel free to choose the definition that better suits their applications.} Gravitational field lines are curves that are tangent to the direction of gravity along their lengths, and therefore normal to the equipotential surfaces.

Let \(\vb{p}\) and \(\vb{q}\) be two points in space. Point \(\vb{p}\) is considered to \textit{overhang} point \(\vb{q}\) if and only if \(\vb{p}\) has a higher gravitational potential than \(\vb{q}\) and also lies on the same gravitational field line as \(\vb{q}\).

The \textit{planetary surface}, denoted by \(\mathbb{S}\), is defined to consist of all points on a planet that are not overhung by another point on the planet, where gravitational acceleration is nonzero. The planetary surface can be thought of as all parts of a planet that are directly exposed to falling raindrops, provided these raindrops are infinitesimally small and fell along the field lines. Geometrically, \(\mathbb{S}\) is a discontinuous surface that can be represented as the union of many continuous surfaces.

\begin{figure}[H]
\centering
\frame{\includegraphics[width=0.5\columnwidth]{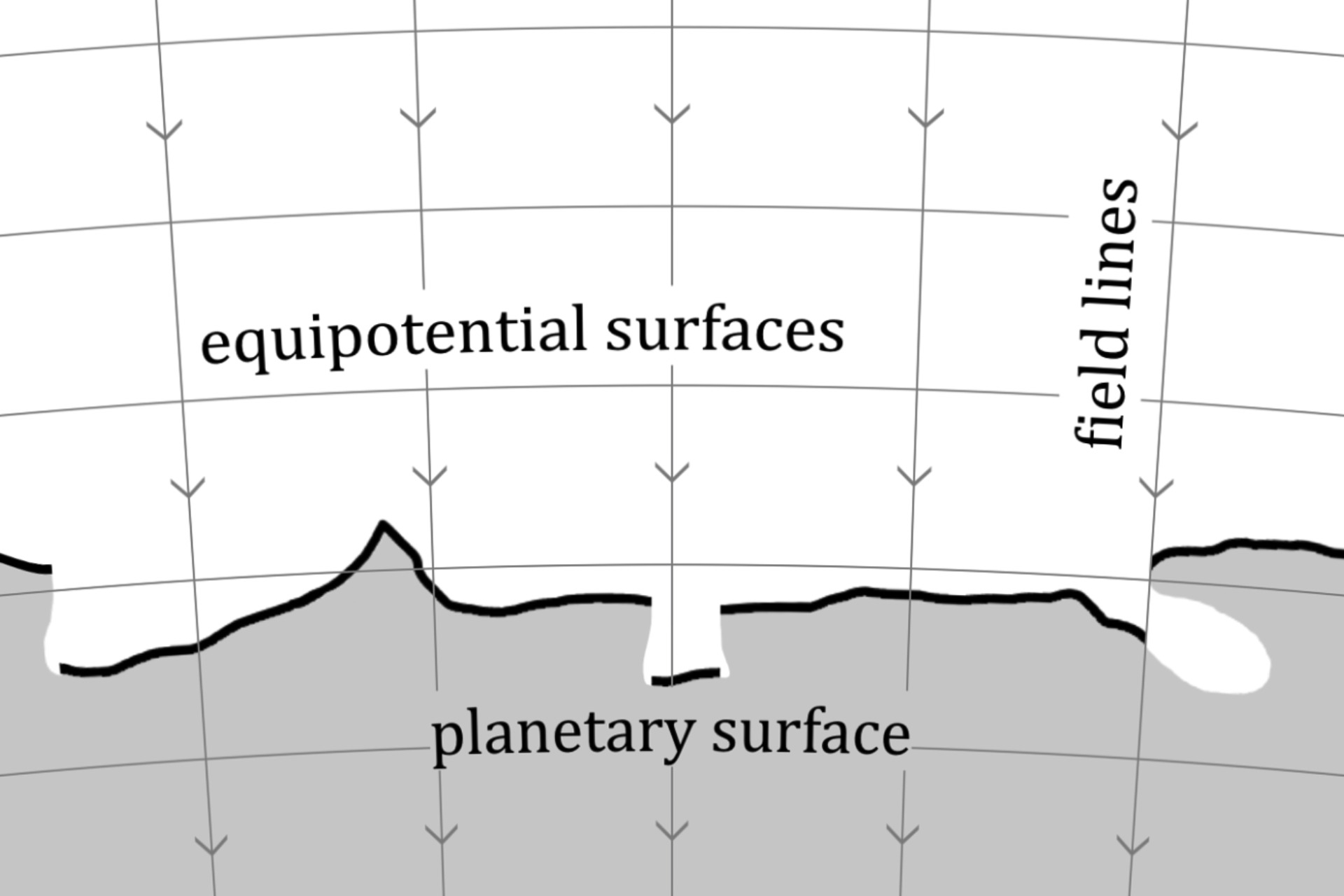}}
\caption{
Points on the planetary surface \(\mathbb{S}\) are outlined in black.
}
\end{figure}

\section{Datumless Surface Area}

Let \(S\) be a subset of the planetary surface (\(S \subseteq \mathbb{S}\)) denoting the region with a surface area of interest.\footnote{Any shape in space (for instance, a region on a reference ellipsoid) can be projected to a corresponding region on the planetary surface by including in \(S\) every point in \(\mathbb{S}\) whose gravitational field line intersects the shape.} The \textit{datumless surface area} of \(S\) is equal to the flux of unit vectors pointing in the direction of gravity through \(S\).

More formally, datumless surface area is notated as follows: At a particular point in \(S\), let \(\vu{g}\) denote the unit vector pointing in the direction of gravity, and let \(\vu{n}\) denote the unit normal vector to \(S\) that forms a less-than-\ang{90} angle with \(\vu{g}\); in other words, \(\vu{n}\) is the unit normal vector that points into the body of the planet, rather than into outer space.\footnote{Note that \(\vu{n}\) is only necessary for notating flux, and does not need to be computed to find datumless surface area.} The datumless surface area of \(S\), denoted by \(A(S)\), is equal to the following flux value:
\begin{equation}
A(S) = \iint_{S} \vu{g} \vdot d\vb{S}
\end{equation}
where \(d\vb{S} = \vu{n} \, dS\), a notational convention in physics.

\begin{figure}[H]
\centering
\frame{\includegraphics[width=0.5\columnwidth]{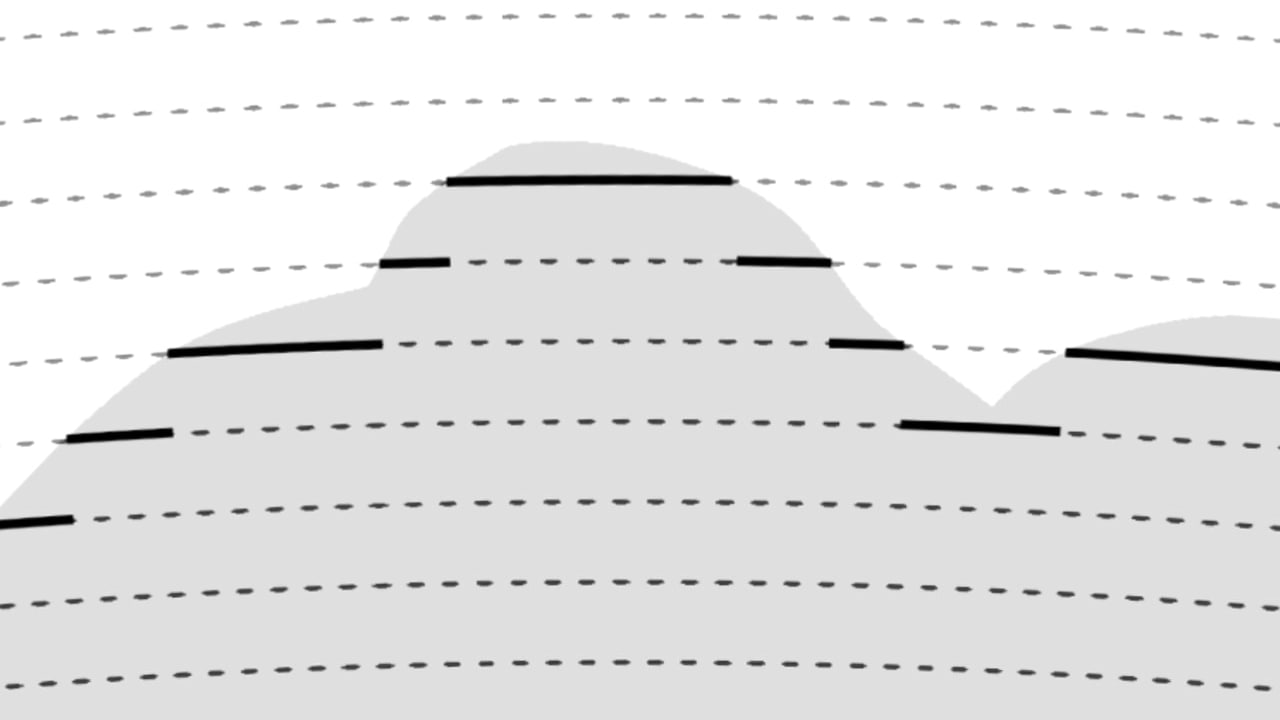}}
\caption{
Datumless surface area can be visualized as the sum of the surface areas of infinitesimally small portions of equipotential surfaces that intersect \(S\).
}
\end{figure}

\begin{figure}[H]
\centering
\frame{\includegraphics[width=0.5\columnwidth]{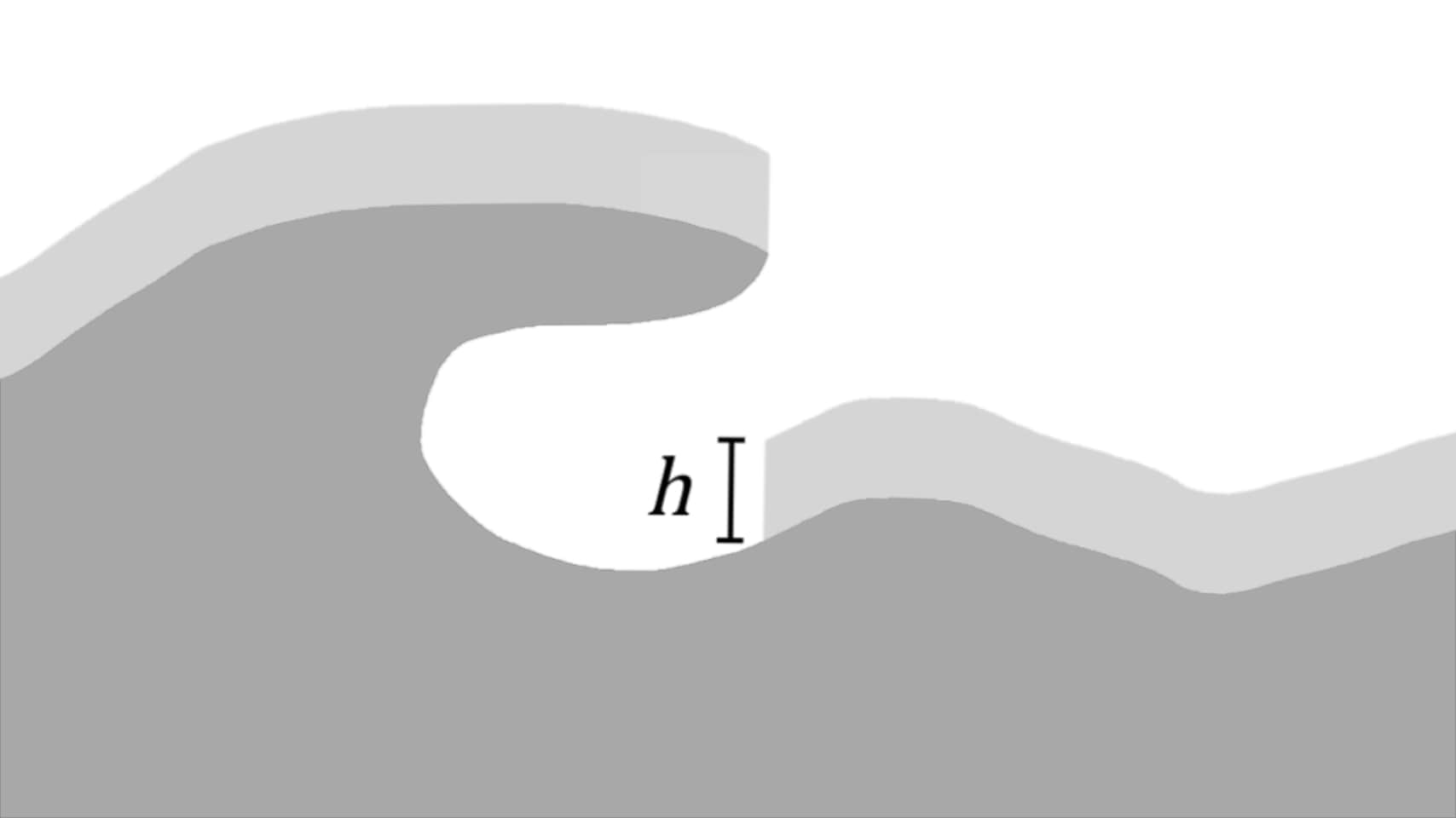}}
\caption{
Alternatively, datumless surface area can be visualized by imagining snow accumulated along the gravitational field lines to height \(h\). Let \(V\) be the total volume of accumulated snow. Datumless surface area equals the limit of \(V / h\) as \(h\) approaches 0.
}
\end{figure}

Datumless surface area describes the horizontal, normal-to-gravity coverage of a landscape, making it useful for measuring land area in the usual bird's-eye-view sense. Vertical cliffs do not contribute to datumless surface area. The same cannot be said about ellipsoid surface area, as the ellipsoid normal may deviate from the local direction of gravity, especially on irregularly-shaped asteroids and comets. On such objects, datumless surface area continues to provide perceptually accurate measurements, doing so in a way that is inherent to the object's physical properties. On fairly round planets such as Earth and Mars, datumless surface area yields very similar values as ellipsoid surface area.
\begin{table}[H]
\centering
\caption{Datumless surface area versus ellipsoid surface area for Earth, Mars, and Vesta. All computations are made in Google Earth Engine \parencite{google-earth-engine}.}
\def\arraystretch{1.2}
\resizebox{\columnwidth}{!}{%
\begin{tabular}{|l|l|l|l|l|l|}
\hline
Object & Datumless SA & Ellipsoid SA & Ellipsoid & Equatorial Axis & Flattening \\ \hline
Earth & \SI{510.084e6}{\kilo\meter\squared} & \SI{510.066e6}{\kilo\meter\squared} & WGS 84 \parencite{earth-ellipsoid} & \SI{6378137}{\m} & 1/298.257224 \\
Mars & \SI{144.653e6}{\kilo\meter\squared} & \SI{144.372e6}{\kilo\meter\squared} & Ardalan et al. \parencite{mars-ellipsoid} & \SI{3395428}{\meter} & 1/191.291713 \\
Vesta & \SI{8.68e5}{\kilo\meter\squared} & \SI{8.66e5}{\kilo\meter\squared} & Konopliv et al. \parencite{vesta-ellipsoid-geoid} & \SI{281000}{\meter} & 1/5.109091 \\ \hline
\end{tabular}
}
\end{table}

\begin{figure}[H]
\centering
\includegraphics[width=\columnwidth]{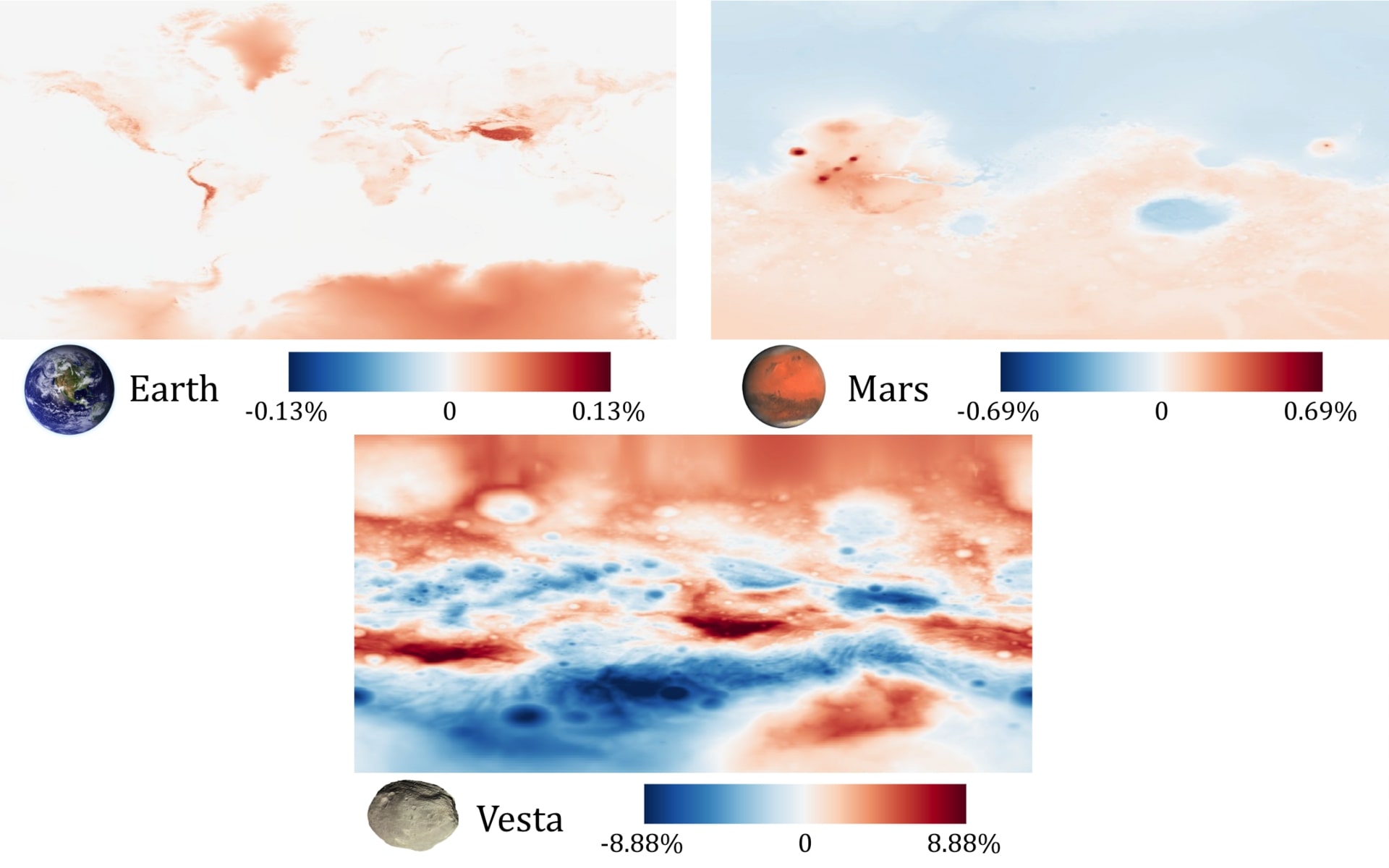}
\caption{
These maps show the deviation of local datumless surface area from local ellipsoid surface area on the surfaces of Earth \parencite{earth-image}, Mars \parencite{mars-image}, and Vesta \parencite{vesta-image}. Red means local datumless surface area is greater than local ellipsoid surface area, and blue means local datumless surface area is less than local datumless surface area. For instance, on the map of Earth, the datumless surface area of the Tibetan Plateau exceeds its ellipsoid surface area, primarily due to its high elevation. The color palette is adapted from ColorBrewer \parencite{colorbrewer}.
}
\end{figure}

\section{Datumless Path Length}

Two points \(\vb{p}\) and \(\vb{q}\) on the planetary surface \(\mathbb{S}\) are considered \textit{neighboring} points if either \(\vb{p}\) and \(\vb{q}\) are infinitesimally close to each other, or there exists a point \(\vb{p}'\) that lies on the same gravitational field line as \(\vb{p}\) and that is infinitesimally close to \(\vb{q}\).

A \textit{path} from point \(\vb{a}\) to point \(\vb{b}\) on the planetary surface, denoted by \(C\), refers to a sequence of points that starts at \(\vb{a}\), ends at \(\vb{b}\), and contains a theoretically infinite number of points in-between, such that all adjacent pairs of points in the sequence are neighboring, i.e., \(C = (\vb{a}, \vb{c}_1, \vb{c}_2, \ldots, \vb{c}_\infty, \vb{b})\). 
A continuous curve in space (such as a flight trajectory or a path on a reference ellipsoid) can be projected to a corresponding path on the planetary surface by mapping every point along such a curve to the point in \(\mathbb{S}\) that lies on its gravitational field line.

Note that while \(C\) may be a discontinuous curve due to the discontinuous nature of the planetary surface, it can be represented as a union of many continuous curves, i.e., \(C = \bigcup_i C_i\). The \textit{datumless path length} of \(C\), denoted by \(L(C)\), is equal to the following:
\begin{equation}
L(C) = \sum_i \int_{C_i} \abs{\,\vu{g} \times d\vb{r}_i}
\end{equation}
where \(d\vb{r}_i\) is the parametrization of the continuous curve \(C_i\).\footnote{Note that \(d\vb{r}_i\) does not actually need to be computed to find datumless path length.}

Datumless path length describes the horizontal, normal-to-gravity coverage of a path on the planetary surface. As such, movements in the direction of gravity do not contribute to datumless path length. The same cannot be said for path length on an ellipsoid, as the ellipsoid normal may deviate from the local direction of gravity, especially on many asteroids and comets. On fairly round planets, the values of datumless path length are very similar to those of ellipsoid path length.

The minimal datumless path length from \(\vb{p}\) to \(\vb{q}\) on the planetary surface can be called the \textit{datumless geodesic distance} between \(\vb{p}\) and \(\vb{q}\).

\section{Datumless Mean Value}

Let \(f\) be a \textit{surface function}, i.e., a function that maps points in \(\mathbb{S}\) to numerical values. Examples of such functions include surface temperature, elevation, and gravitational acceleration.\footnote{Note that \(f\) could also be a non-surface attribute (for instance, temperature at a certain altitude) projected onto the planetary surface along the gravitational field lines.} Used to average \(f\) over a region, the \textit{datumless mean value} of \(f\) over region \(S\), denoted by \(\Bar{f}(S)\), is equal to the following expression derived from datumless surface area:
\begin{equation}
\Bar{f}(S) = \frac{1}{A(S)} \, \iint_{S} f \, \vu{g} \vdot d\vb{S}
\end{equation}
Similarly, the datumless mean value of \(f\) along path \(C\), denoted by \(\Bar{f}(C)\), is the following:
\begin{equation}
\Bar{f}(C) = \frac{1}{L(C)} \, \sum_i \int_{C_i} f \, \abs{\vu{g} \times d\vb{r}_i}
\end{equation}

\section{Datumless Statistical Parameters}

Any statistical parameter of a surface function \(f\) within a region or path (such as median, quantiles, standard deviation, etc.) can be defined without a datum. To do so, let \(I_{f \leq x}(\vb{p})\) be an indicator function that equals 1 if \(f\) is less than or equal to \(x\) at point \(\vb{p}\), and 0 otherwise. The \textit{datumless percentile} of value \(x\) for function \(f\), denoted by \(P(f \leq x)\), is equal to the datumless mean value of \(I_{f \leq x}\) on the region or path:
\begin{equation}
P(f \leq x) = \bar{I}_{f \leq x}
\end{equation}

Knowing the percentiles of all values of \(x\) that a function can assume within a region or path, a distribution of the values of \(f\) within the region or path can be constructed. From such a distribution, any statistical parameter can be derived.

\section{DSA-Gravity-Mass Identity}

Let \(S\) be a subset of the planetary surface (\(S \subseteq \mathbb{S}\)) denoting the region with a surface area of interest. Let \(A\) denote the datumless surface area (DSA) of \(S\), and let \(\bar{g}\) denote the datumless mean value of gravitational acceleration on \(S\). Let \(M\) denote the mass of all matter that is overhung by some point in \(S\).

\begin{figure}[H]
\centering
\includegraphics[width=0.5\columnwidth]{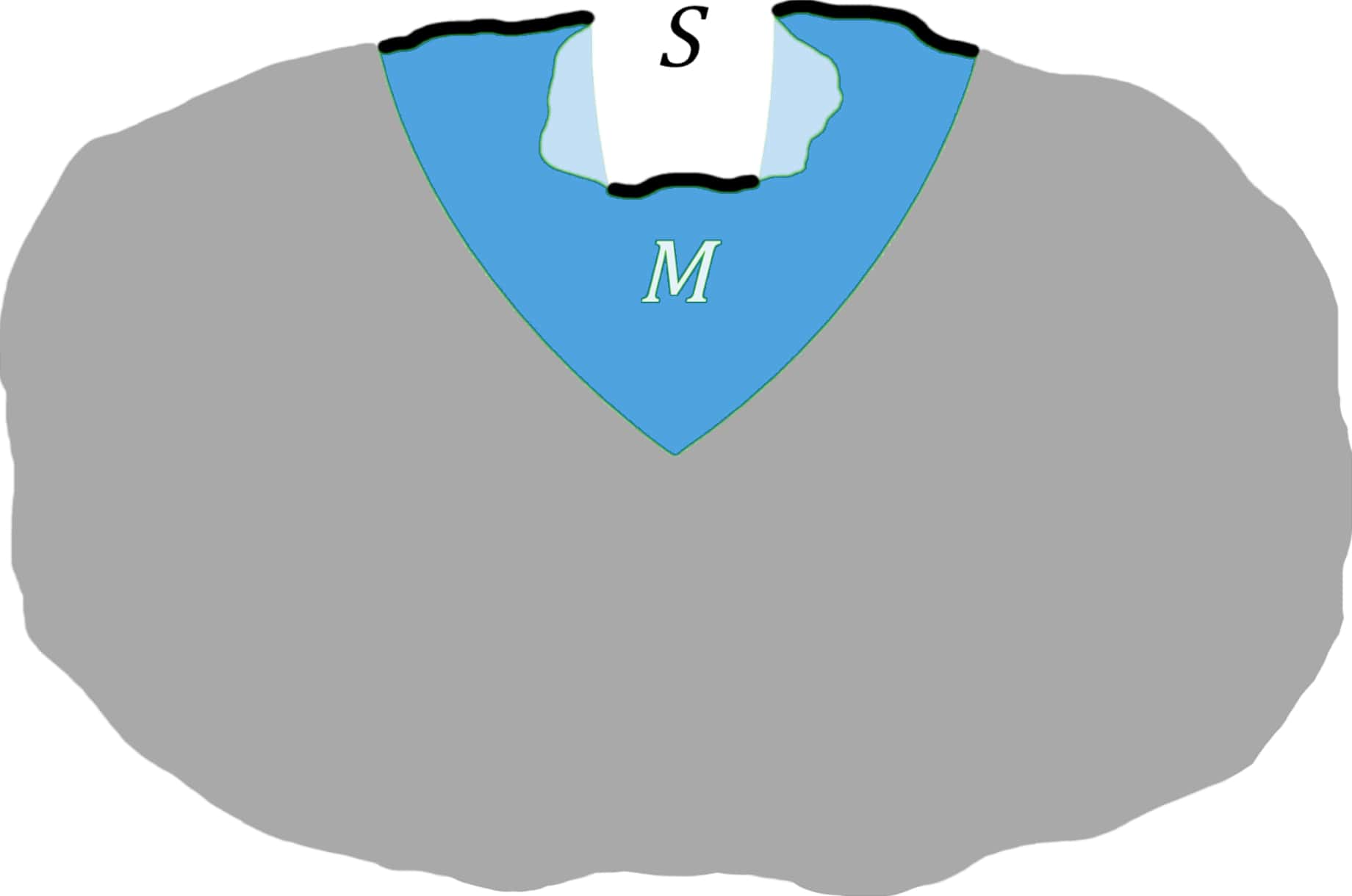}
\caption{
Diagram of the mass \(M\) (shaded in blue) that is overhung by \(S\) (outlined in black).
}
\end{figure}

From Gauss's law, the product of \(\bar{g}\) and \(A\) (which is equal to the flux of gravitational acceleration through \(S\)) is directly proportional to \(M\):\,\footnote{For the theorems in this section to hold, gravity must be defined as the fundamental force of gravitation, rather than gravity in the geophysicist's sense, i.e., gravitation plus centrifugal force.}
\begin{equation}
\bar{g} A = 4\pi GM
\end{equation}
where \(G\) is the gravitational constant, not to be confused with gravitational acceleration.

Let \(g_\text{max}\) and \(g_\text{min}\) denote the maximum and minimum gravitational acceleration found anywhere on \(S\). By virtue of being a mean value, \(\bar{g}\) is no less than \(g_\text{min}\) and no greater than \(\bar{g}_\text{max}\). Therefore, the datumless surface area of \(S\) is finitely bounded by this inequality:
\begin{equation}
\frac{4\pi GM}{g_\text{max}} \leq A \leq \frac{4\pi GM}{g_\text{min}}
\end{equation}

In isolated systems, the mass that is overhung by the entire planetary surface \(\mathbb{S}\) is equal to the total mass of the system.

\section{Computing the Datumless Measures}

The datumless measures can be computed in GIS software using a digital elevation model and a gravity model such as a geoid. This is done by finding how much the datumless measures deviate from corresponding measures on a reference ellipsoid. A positive deviation results from elevation being above the ellipsoid, or offsets in the local direction of gravity from the ellipsoid normal. A negative deviation results from elevation being below the ellipsoid.

Deviations as a result of elevation are represented by the \textit{elevation correction}, denoted by \(k_\text{elev}\). The value of \(k_\text{elev}\) varies at different points on the planetary surface, and is usually slightly above or below 1. At a particular point, let \(x\), \(y\), and \(z\) denote the ECEF coordinates at that point, as converted from latitude, longitude, and height above the ellipsoid. With \(a\) denoting the length of the equatorial axis of the reference ellipsoid, and \(b\) denoting the length of the polar axis, the value of \(k_\text{elev}\) at that point is equal to the following:
\begin{equation}
k_\text{elev} = \sqrt{\frac{x^2 + y^2}{a^2} + \frac{z^2}{b^2}}
\end{equation}
Measurements in this paper use the ALOS World 3D-30m \parencite{earth-dem}, MGS MOLA 463m v2 \parencite{mars-dem}, and Dawn FC HAMO Global 93m v1 \parencite{vesta-dem} digital surface models on Earth, Mars, and Vesta, respectively.

Deviations as a result of offsets in the direction of gravity are represented by the \textit{gravity correction}, denoted by \(k_\text{grav}\). The value of \(k_\text{grav}\) is at least 1 anywhere on the planetary surface, and is usually much closer to 1 than the elevation correction. At a particular point, let \(\theta\) denote the angle between the direction of gravity and the inward-pointing ellipsoid normal at that point, which can be approximated by taking the slope of a geoid height model in GIS software. The value of \(k_\text{grav}\) at that point is equal to the following:
\begin{equation}
k_\text{grav} = \sec(\theta)
\end{equation}
Measurements in this paper use the EGM2008 \parencite{earth-geoid-1, earth-geoid-2}, MOLA MEGDR \parencite{mars-geoid}, and Dawn v2 \parencite{vesta-ellipsoid-geoid} geoid models for Earth, Mars, and Vesta, respectively. The MOLA MEGDR geoid is derived using Ames Stereo Pipeline \parencite{ames-stereo-pipeline}.

At a point on the planetary surface, the \textit{total correction}, simply denoted by \(k\), is equal to the product of the elevation correction and gravity correction at that point:\,\footnote{On the maps showing the deviation of local datumless surface area from local ellipsoid surface area on various planets, the value of \(k - 1\) was displayed.}
\begin{equation}
k = k_e \cdot k_g
\end{equation}

Datumless surface area and datumless path length can be approximated by multiplying ellipsoid surface area or ellipsoid path length with the ellipsoid mean value of \(k\):
\begin{equation}
A(S) \approx A_\text{ellip}(S) * \bar{k}_\text{ellip}(S)
\end{equation}
\begin{equation}
L(C) \approx L_\text{ellip}(C) * \bar{k}_\text{ellip}(C)
\end{equation}
where \(A_\text{ellip}\), \(L_\text{ellip}\), and \(\bar{k}_\text{ellip}\) denote ellipsoid surface area, ellipsoid path length, and ellipsoid mean value of \(k\), respectively.

The datumless mean value of a function \(f\) over a region or path can be approximated by taking a weighted ellipsoidal average of the product of \(f\) and \(k\), weighted by \(k\):
\begin{equation}
\bar{f} \approx \frac{\overline{fk}_\text{ellip}}{\bar{k}_\text{ellip}}
\end{equation}
where \(\overline{fk}_\text{ellip}\) denotes the ellipsoid mean value of the product of \(f\) and \(k_\text{ellip}\).

\section{Acknowledgements}

I am grateful to Tony Wang and Barkotel Zemenu for helping to clarify the mathematical notation used in this paper. I would also like to thank the Yale Undergraduate Research Association for providing the opportunity to present several key ideas in this paper.

\section{References}

\printbibliography[heading=none]

\end{document}